\documentstyle[12pt]{article}
\columnwidth\textwidth

\newcommand{\be}{\begin{equation}}
\newcommand{\ee}{\end{equation}}
\newcommand{\bq}{\begin{eqnarray}}
\newcommand{\eq}{\end{eqnarray}}

\def\g{\gamma}

\def\to{\rightarrow}

\def\be{\begin{equation}}
\def\ee{\end{equation}}
\def\bq{\begin{eqnarray}}
\def\eq{\end{eqnarray}}

\begin{document}
\thispagestyle{empty}
\setcounter{page}{0}
\begin{flushleft}
ZU-TH 8/95\\
LMU-06/95\\
June 1995
\end{flushleft}
\vspace*{\fill}
\begin{center}
{\Large\bf
Calculation of Long-Distance Effects \\
in Exclusive Weak Radiative Decays of B-Mesons $^*$}\\
\vspace{2em}

\large
{\bf A. Khodjamirian$^{a,\dagger,1}$,
G. Stoll$^b$, D. Wyler$^b$ }\\

\vspace{4em}
{\small
$^a$Sektion Physik der Universit\"at M\"unchen\\
D-80333 M\"unchen , Germany } \\
\vspace{2em}
{\small $^b$Universit\"at Z\"urich, 8057
  Z\"urich, Switzerland} \vspace*{0.5cm}

\end{center}
\vspace*{\fill}

\begin{abstract}
We calculate the contribution of the weak annihilation to the
$B\rightarrow \rho\gamma$  decay amplitude by means of QCD sum rules
using the photon light-cone wave function. We find that this
long-distance contribution amounts to about 10\% of the leading
short-distance effect in $B^-\rightarrow \rho^-\gamma$.
On the other hand, weak annihilation is the dominant
source of the corresponding $D$ meson decays and according to our
estimates, yields branching ratios of $O(10^{-4})$ for
$D^0\rightarrow \bar{K}^{*0}\gamma$, $O(10^{-5})$ for
$D_s\rightarrow \rho^+\gamma$, and $O(10^{-6})$ for
$D^-\rightarrow \rho^-\gamma$ and for $D^0\rightarrow \rho^0\gamma$.
\end{abstract}

\vspace*{\fill}

\begin{flushleft}
\noindent$^\dagger${\small on leave from
Yerevan Physics Institute, 375036 Yerevan, Armenia} \\
\noindent$^*${\it partially supported by
    Schweizerischer Nationalfonds.} \\
\noindent{  $^1$ supported
by the German Federal Ministry for Research and
Technology under contract No. 05 6MU93P\\ }

\end{flushleft}


\newpage

1. Rare decays of $B$-mesons, such as the recently observed
processes $B\rightarrow K^*\gamma$ and $ B \to X_s \g$
\cite{COR1,COR2} are becoming an
important tool for studying new forces beyond the standard
model \cite{miko}. Indeed, many
authors have investigated the effects of multi-Higgs models,
supersymmetric theories, left-right symmetric models, etc. on
this process \cite{Hewett}. The interest in these decays stems from
the fact that they occur only through loops and are
therefore particularly sensitive to ``new physics''.

Besides the $b \to s$ transition,
there are also $b \to d$ processes which are suppressed by
the ratio  $V_{td}/V_{ts} \approx 0.1$ of CKM matrix elements.
Within the standard model the rates are then estimated to be
about $10^{-5}$ and $10^{-6}$ for $b \to d\gamma$ and
$B\rightarrow \rho\gamma$, respectively.
This mode is sensitive to other CKM matrix elements (and
possibly to other new physics) and may
show large CP-violation \cite{gre3}.

The "inclusive" processes like $b \to s \gamma $ can be usually evaluated
on the quark level within perturbative methods \footnote{An
exception is the "long distance" part of the
penguin mechanism (see below)},
but are more difficult experimentally. Exclusive processes,
on the other hand are
more easily observable, but require difficult non-perturbative
calculations of matrix elements in order to yield useful
results.

It is generally believed that the short distance penguin
mechanism dominates the exclusive decays. The corresponding
matrix element has been calculated with various methods
\cite{gre3,ali3} which yield similar results; this indicates that
the related uncertainties are
presumably under
control. Besides the matrix element of the penguin operator,
there are other long-distance effects which must
be analyzed. Several authors have recently investigated
these effects
\cite{chengn,golpa,desp3,atwo3,burdgo,soares}
in a phenomenological way. The crude estimates
show that
the long distance contributions can indeed
be small. However, the results obtained vary
enormously
and can  give at best an order of magnitude
estimate. Clearly, a more reliable calculation is desired.

We may distinguish two long distance effects. First, there is a
penguin type mechanism where the low momentum
part of the usual penguin loop must be included.
The above mentioned estimates so far indicate that it is at
most several percent
of the short distance result both for exclusive
and inclusive processes. In these estimates, the radiative
transition $B \rightarrow M \gamma $
is modelled by a sum over the processes
$B\rightarrow M V^*$  followed by the transition
$V^* \rightarrow \gamma $
where $V^*$ is a virtual vector meson with appropriate quantum numbers.
The problems with this
picture are obvious: the effective couplings (for the primary
vertex and for the transition of the resonance into a photon) are
only measured at the mass of the resonance and must be
scaled to zero. Then, some information about the
various structures of the primary vertex must exist,
since only the transversal part can contribute to the
photon transition. In the existing literature, one
or the other of these problems is not addressed,
 but there is generally a strong
suppression. For the particular case of $B\rightarrow K^* \gamma$
this suppression however, crucially  depends
on the model assumptions about the
share of the factorizable contribution.
If factorization dominates, we can combine the results of
refs. \cite{soares} and
\cite{desp3} to predict a negligible effect for the $J/\Psi$
resonance. But only a more systematic calculation
combined with direct estimate of nonfactorizable effects
can clarify the role of this mechanism.

Then, there is a completely different class of
diagrams which do not involve loops of
heavy (new) quarks but just the ordinary four fermion interaction.
In these graphs, the 'spectator' participates in the weak
annihilation; we therefore call them weak annihilation (WA).
Because of the CKM matrix elements, this mechanism is negligible
for $B\rightarrow K^*\gamma$ but important for other exclusive
radiative decays.  For definiteness, we will consider it for the
decay $B\rightarrow \rho\gamma$, although other decays can
be treated in similar ways and will be discussed
briefly at the end. Fig. 1 represents schematically
the mechanism.
A perturbative evaluation of
the diagrams in Fig.1 \cite{chengn,atwo3}
is rather uncertain,
due to the almost on-shell propagation of the light
quark which implies
the use of a the poorly understood 'constituent'
quark mass. The
 phenomenological description via the chain $B\rightarrow \rho\rho*
\rightarrow \rho\gamma$ \cite{atwo3}
suffers from the unknown $B\rho\rho$ coupling
and the problems connected to the off-shellness of the $\rho$.

In this paper we show that existing QCD sum-rule
technology allows us to provide
reliable predictions for the long distance
effects. We will discuss in this short note only
the WA effects which are more uncertain in the
phenomenological approach and which are important
by themselves to a large class of transitions. We
will only be interested in the leading effects;
more accurate and further results will be presented in
a future publication; we will asses here some of the remaining
uncertainties.
\vspace*{2.0cm}

2.  The relevant effective Hamiltonian for
$B\rightarrow \rho\gamma$ consists of two operators \cite{opera}:
\be
\label{H}
{\cal H}_W= \frac{G}{\sqrt{2}}V_{ub}V^*_{ud}\{(
c_1(\bar{d}L_\mu u)(\bar{u}L^\mu b)+
c_2(\bar{u}L_\mu u)(\bar{d}L^\mu b)\}~,
\label{Hw}
\end{equation}
where $L_\mu = \gamma_\mu(1-\gamma_5)$,
and $c_1$ and $c_2$ are the Wilson coefficients.
For the decay  $B^-\rightarrow \rho^-\gamma$
the combination
\be
\label{effop}
{\cal H}_W= \frac{G}{\sqrt{2}}V_{ub}V^*_{ud}
a_1(\bar{d}L_\mu u)(\bar{u}L^\mu b) ~,
\ee
with $a_1 = c_1 + c_2/3$ enters. Similarly, the
combination $a_2 = c_1/3 + c_2$
multiplies the corresponding operator for $B^0\rightarrow \rho^0\gamma$.
Following the phenomenological ansatz \cite{bsw}, the coefficients
$a_1$, $a_2$ are extracted from two-body
non-leptonic decays using factorization.  Recent fits
\cite{CLEO3} indicate a considerable departure of $a_2$ from
its short distance value (see, however, the recent discussion
in \cite{Buras}). The reason for that departure
may be connected with nonfactorizable contributions to the hadronic
matrix element,
which in the decay $B^-\rightarrow \rho^-\gamma$ come from the
color octet operator
$(\bar{d}L_\mu \frac{\lambda^a}2 u)
(\bar{u}L^\mu \frac{\lambda^a}2 b)$ with coefficient $2c_2$.

The actual size of nonfactorizable effects is
under discussion
at present. First direct estimates \cite{BS},
\cite{KR},\cite{Halp} by QCD sum rule methods
indicate that nonfactorizable effects in $B$ decays
are quite process dependent.
Moreover, it is not clear if one can use
the same value of the effective parameters $a_{1,2}$
for nonleptonic and radiative decays, processes with
different physical final states.
Nevertheless,
since the decay $B^-\rightarrow \rho^-\gamma$ depends on
the parameter $a_1$
which is near its perturbative value $a_1 \sim 1 $,
we assume for the present
work that factorization is sufficient. At the same time one should keep
in mind that the
result for  $B^0\rightarrow \rho^0\gamma$
which depends on $a_2$
will only be approximate.

Turning to more detailed analysis of the WA mechanism we first of all
neglect the photon emission from the final state quarks
invoking the well known helicity arguments:
being factorized, the matrix element
of (\ref{effop}) is in this case proportional to the light quark masses:
$
\langle \rho^- \mid H_W \mid B^-\rangle _\gamma \sim
 p_B^\mu f_B
\langle \rho^- \mid (\bar{d} L^\mu u) \mid 0 \rangle _\gamma
\sim O(m_u,m_d).
$
Note however
that non-factorizable interactions such as gluon exchanges between
final and initial quarks lift this
suppression.

Thus we are left with two diagrams in Figs. 1a,b
where the photon is emitted from
the initial $b$ and $u$ quark lines. The corresponding matrix
element can be written as
\be
\langle \rho^- \mid H_W \mid B^-\rangle _\gamma =
\frac{G}{\sqrt{2}}V_{ub}V^*_{ud}a_1f_\rho m_\rho \epsilon^\mu_\rho
\langle 0 \mid (\bar{u}L_\mu b) \mid B^- \rangle_\gamma,
\label{matrel}
\end{equation}
where we used
$
\langle \rho^- \mid (\bar{d}L^\mu u) \mid 0 \rangle
=f_\rho m_\rho \epsilon_\rho^\mu,
$
denoting by $f_{\rho}$ and $\epsilon_\rho$ the decay constant
and the polarization vector of the charged $\rho$-meson.
It then remains to calculate the matrix element
\be
\langle 0 \mid (\bar{u}L^\mu b) \mid B^- \rangle_\gamma=
-A_{PC}\varepsilon_{\mu\tau\lambda\sigma}p^\tau \epsilon^\lambda
q^\sigma+iA_{PV}[q_\mu(\epsilon \cdot p) -\epsilon_\mu (p \cdot q) ]~,
\label{brhogamma}
\ee
which describes the
annihilation of $B^-$ into the current $\bar{u}L^\mu b$ with
momentum $p$  accompanied by
the emission of a real photon with momentum $q$ and polarization
vector $\epsilon$.
We have written explicitly
decomposition of this matrix element into two gauge invariant
(with respect to the electromagnetic field) terms
$A_{PC}( A_{PV} ) $
corresponding to parity conserving ( parity-violating)
$B \rightarrow \rho \gamma $ decay amplitude.
We note
that for the short distance penguin mechanism
one has $A_{PC} = A_{PV}$ if the
light quark mass is neglected ( see e.g. \cite{gre3}).
In principle, polarization experiments
could distinguish the two amplitudes and thus determine to
what extent this equality is valid.
In fact, it turned out that the WA mechanism considered here
does not respect it, but the deviation is due to nonleading
contributions and consequently small.

As noted, we propose to use QCD sum rules in order
to calculate the matrix
element (\ref{brhogamma}).
Since the photon emission from
the light quark takes place
at large distances, the use of standard QCD sum rules \cite{SVZ}
based on the local operator
product (OPE) expansion
is not sufficient. Rather, one should use a light-cone
expansion which is adequate for exclusive processes with light
particles. It will involve the hadronic wave functions on the light-cone
which encode the photon emission by a quark-antiquark pair
at light-like separation in close analogy to the well known pion wave
function \cite{exclusive}. The photon light-cone wave function
was introduced in ref. \cite{bbk}
for calculating  the amplitude of weak radiative decay
$\Sigma \rightarrow \rho \gamma$ and used later in
ref. \cite{bf} to evaluate the nucleon magnetic moments.
\vspace*{2.0cm}

3. We now proceed to calculate the correlation function
\be
\label{corr}
\Pi_\mu (p,q)=
i\int d^4xe^{ipx} \langle 0|T\{ \bar{u}(x)L_\mu b(x),
\bar{b}(0)i\gamma_5 u(0)\} |0\rangle_{F(q)} ~.
\ee
in the external electromagnetic field
\be
F_{\alpha\beta}(q,x)
= i(\epsilon_\beta q_\alpha - \epsilon_\alpha q_\beta)
e^{iqx}
\label{Fem}
\ee
with momentum $q$ and polarization vector $\epsilon$.
The function $\Pi_{\mu}$ can be decomposed into two parts
corresponding to parity
conserving and a parity violating part in (\ref{brhogamma}):
\be
\Pi_\mu=
\Pi_{PC}i\varepsilon_{\mu\tau\lambda\rho}p^\tau \epsilon^\lambda q^\rho
+\Pi_{PV}[q_\mu(\epsilon \cdot p) -\epsilon_\mu (p \cdot q) ]
\label{decom}
\ee
In the region $(p+q)^2 < 0$  and at $p^2=m_{\rho}^2 << m_{b}^2$, the
heavy $b$-quark is far off-shell.
In particular, photon emission from the heavy $b$-quark
takes place perturbatively.
The accompanying
light-quark propagator may then be described by the local OPE
containing as a first approximation the free propagation
and in the next orders,
interaction with quark and quark-gluon vacuum condensates.
The  corresponding contributions to the correlation function are
depicted in Figs. 2a, b and c, respectively.
As far as photon emission from the light $u$-quark is concerned,
after contracting the $b$-quark
line, one is left with matrix element
\be
\Pi_\mu(p,q)=i\int \frac{
d^4x\,d^4k}{(2\pi )^4(m_b^2-k^2)}
e^{i(p-k)x}
\langle 0|\bar{u}(x)L_\mu(m_b+\not k)\gamma_5u(0)|0\rangle_{F(q)}.
\label{matr}
\ee
The diagram Fig. 2d describes only the short-distance part
of this matrix element corresponding to the photon emission from
freely propagating $u$ quark.

To take into account the long-distance part one needs
to introduce additional nonperturbative parameters, describing
the light-quark propagation in the presence of external
electromagnetic field
which is schematically shown in diagram of Fig. 2e.
In the case of constant
electromagnetic field ($q\rightarrow 0$ )
one still can use the local OPE i.e. expand
(\ref{matr}) around $x=0$ in the
external field \cite{IoffeSmilga,BalYung}.
Physically, the most important
parameter emerging in this expansion is
the so called induced quark condensate
\be
\langle 0|\bar{q}\sigma_{\alpha\beta}q|0\rangle_{F}=
e_q \langle\bar{q}q\rangle \chi F_{\alpha\beta}
\label{chi}
\ee
where $e_q$ is the quark electric charge,
$\langle\bar{q}q\rangle$ is the quark condensate density
and parameter $\chi$ has physical meaning of
magnetic susceptibility of the quark condensate.

However, here we deal with essentially large photon
momenta of the order
of heavy quark mass.
In this kinematical configuration the use of local OPE will
lead to unmanageable infinite
series of condensates for the expansion of (\ref{matr})
(see a recent detailed discussion in \cite{BBKR}).
The adequate approach is to use the expansion
near the light-cone $x^2=0$ introducing
matrix elements of nonlocal operators.
The leading twist two contribution \cite{bbk,bf} emerges from the
expansion of the operator $\bar{u}(x)\sigma_{\alpha\beta}u(0)$ and
the corresponding part of the matrix element in the external
photon field can be parametrized as\footnote{ the path-ordered
exponential gauge factors for both gluon and photon fields are not
shown for brevity}
\be
\langle 0|\bar{u}(x)\sigma_{\alpha\beta}u(0)|0\rangle_{F(q)}=
e_u \langle \bar{u}u\rangle \int^1_0 du \varphi(u)F_{\alpha\beta}(ux)~.
\label{wf}
\ee
Here, the function $\varphi(u)$ has the meaning of
the photon wave function in terms of its
quark-antiquark constituents
and may be interpreted as the distribution in the fraction of
light-cone momentum $q_0 +q_3$ of the photon carried by a quark.
The asymptotic form of this wave function is known
\cite{exclusive,bbk}:
\be
\varphi(u) = 6\chi u(1-u)~,
\label{as}
\ee
where the appearance of the magnetic susceptibility $\chi$ is
evident from taking local limit of (\ref{wf}) and comparing with
the definition (\ref{chi}).
Moreover, according to the analysis carried out in
ref. \cite{bbk},
non-asymptotic effects in $\varphi(u)$ and higher twist (twist 4)
contributions to (\ref{wf}) are small, contrary to
the case of the pion wave functions. Therefore, in what follows we
will  use the asymptotic wave function (\ref{as}). A rough
estimate of the twist four contribution carried out below will
allow us to neglect all higher twist effects.

To proceed, we retain in eq. (\ref{matr}) the parts
containing $\bar{u}(x)\sigma_{\mu\nu}u(0)$, insert eq. (\ref{wf}),
and after integration over $x$ and $k$ obtain
the leading twist 2 contribution to the correlation
function (\ref{corr}) corresponding to the
long distance photon emission from u-quark and depicted in Fig. 2e:
\be
\Pi_{PC}^{twist2}=\Pi_{PV}^{twist2}=ie_u\langle \bar{u}u
\rangle\int^1_0 du
\frac{\varphi(u)}{m_b^2-(p+uq)^2} ~.
\label{tw2}
\ee

The next important contribution stems from the perturbative
loop diagrams of Figs. 2a and 2d.
Their contribution to the parity conserving invariant amplitude
written in the
form of dispersion integral in variable $(p+q)^2 $ at fixed $p^2 $ is:
\be
\Pi_{PC}^{pert} = \frac{3im_b}{4\pi^2}
\int^\infty_{m_b^2} \frac{ds}{[s-(p+q)^2](s-p^2)}
\left[ (e_u-e_b)\frac{s-m_b^2}{s}+e_bln\frac{s}{m_b^2}\right]~,
\label{loops}
\ee
where the pieces proportional to $e_b$ and $e_u$
correspond to the diagrams Fig. 2a and 2d,
respectively.
The contribution of the loop diagrams to the parity violating amplitude
is:
$$
\Pi_{PV}^{pert} = \frac{3im_b}{4\pi^2}
\int^\infty_{m_b^2} \frac{ds}{[s-(p+q)^2](s-p^2)^2}
\Bigg[ \Bigg\{ e_u\left( 2m_b^2-s- \frac{p^2m_b^2}{s} \right)
$$
\be
-e_b\left( 2m_b^2-s- \frac{p^2(m_b^2-2s)}{s} \right) \Bigg\}
\frac{s-m_b^2}s
+e_b(2m_b^2-s+p^2)ln\frac{s}{m_b^2}\Bigg]~.
\label{loopsPV}
\ee
We see that the amplitudes $\Pi_{PC}$ and $\Pi_{PV}$ are
indeed different.
To assess the importance of nonperturbative effects
in the case when photon is emitted from the heavy quark we have also
calculated the contributions of quark condensate corresponding
to Fig. 2b with the result
\be
\Pi_{PC}^{\langle\bar{q}q\rangle}=
\Pi_{PV}^{\langle\bar{q}q\rangle}=-ie_b\frac{\langle \bar{u}u\rangle}
{(m_b^2-p^2)(m_b^2-(p+q)^2)}~,
\label{cond}
\ee
In calculating the perturbative and quark condensate
contributions, the contact terms violating
gauge invariance with respect to
the photon field should be properly taken into account. If
one carries out the calculation in the fixed point gauge
for the photon field, the origin must be chosen carefully.
The quark-gluon condensate contribution (Fig. 2c) and
higher-twist contributions corresponding to three-particle
photon wave function (Fig. 2f) are neglected in the present
calculation.

To understand the level of accuracy
for the most important contribution
containing the photon wave function we need some estimate of
nonleading higher twist contributions neglected in the matrix element
in eq. (\ref{wf}).
We parametrize the next-to-leading twist 4 contribution in
the expansion of this matrix element as :
\bq
\langle 0|\bar{u}(x)\sigma_{\alpha\beta}u(0)|0\rangle_{F(q)}^{twist 4}
= e_u \langle \bar{u}u\rangle \int^1_0 du \{f_1(u)
F_{\alpha\beta}(ux)x^2
\nonumber
\\
+f_2(u)\left[F_{\alpha\rho}(ux)x_\rho x_\beta
-F_{\beta\rho}(ux)x_\rho x_\alpha  \right]\}~.
\label{wftw4}
\eq
As a result,
\be
\Pi_{PC}^{twist4}=-4ie_u\langle \bar{u}u\rangle\int^1_0 du
\frac{f_1(u)}{[m_b^2-(p+uq)^2]^2}\left[1+\frac{2m_b^2}
{m_b^2-(p+uq)^2} \right]~,
\label{tw4}
\ee
and the corresponding answer for $\Pi_{PV}^{twist4}$
is obtained by replacing $f_1(u)$ in the numerator
of (\ref{tw4}) by the combination $f_1+f_2$.
The only information about distributions
$f_{1,2}$ is available from comparison with
the twist 4 terms in the light-cone sum rules for
the $\Sigma\rightarrow p \gamma$ decay \cite{bbk}
and for the nucleon magnetic moment \cite{bf}
which result in the following relation:
\be
f_1(u)+\frac{f_2(u)}2 = -\frac14\left[(1-u)
-\varphi_4(u)\right] ~.
\label{18}
\ee
The first term in r.h.s. of (\ref{18}) originates from
the expansion of the light quark field
in the electromagnetic field, whereas the wave function
$\varphi_4$ corresponds to analogous expansion in the gluon field
and is related to the three-particle photon wave
functions of twist four via QCD equation of motion. According to
estimates of  \cite{bbk}, $ \varphi_4(u) \simeq 30ku^2(1-u)^2+...$,
with the parameter $k \simeq -0.2 $ and the
nonasymptotic corrections denoted by ellipses and not shown for brevity.
We assume for an order of magnitude estimate that $f_1 =f_2$,
having in mind that the normalization integrals of these two distributions
are equal. We plan to analyze this contribution in more details in future.

4. The QCD sum rule is obtained as usual by equating the
hadronic representation of the correlation function
 $\Pi_{\mu}$ with the result of the QCD calculation, a sum
of all contributions calculated above.
Inserting hadronic states with $B$ meson quantum numbers
into eq.(\ref{corr}), we have the following decomposition for
invariant amplitudes in eq.(\ref{decom}):
\be
\label{hadro}
\Pi_{PC(PV)} = \frac{if_Bm_B^2A_{PC(PV)}}{m_b[m_{B^2}-(p+q)^2]}
+\int^\infty_{s_0}ds\frac{\rho^h_{PC(PV)}(s,p^2)}{s-(p+q)^2}
\ee
where the first term represent the $B$-meson contribution
containing the matrix element
(\ref{brhogamma}) and $\langle B \mid \bar{b}i\gamma_5 u \mid 0
\rangle = f_Bm_B^2/m_b$. The second term in (\ref{hadro})
represents contribution of the higher states in $B$ channel
starting from some effective threshold $s_0$.
We invoke the usual hadron-quark duality and replace the spectral
density $\rho^h_{PC(PV)}$ by the imaginary part
of $\Pi_{PC(PV)}$ calculated in QCD.
Equating now eq. (\ref{hadro}) with  the result of this calculation
given in eqs.
(\ref{tw2})-(\ref{cond}) and applying a Borel
transformation in $(p+q)^2$ to suppress
the higher states we obtain
the desired QCD sum rule for both amplitudes
entering the hadronic matrix element (\ref{brhogamma}).
The result for the parity conserving amplitude is
$$
A_{PC}= \frac{m_b}{f_Bm_B^2}
\Bigg\{\int_\Delta^1\frac{du}{u}
\exp\left[\frac{m_B^2}{M^2}-\frac{m_b^2-p^2(1-u)}{uM^2}\right]
\Bigg[ e_u\langle\bar{u}u\rangle \varphi(u)
$$
$$
+\frac{3m_b}{4\pi^2}\Bigg((e_u-e_b)\frac{(m_b^2-p^2)(1-u)}{m_b^2-p^2(1-u)}
+e_bln\left[\frac{m_b^2-p^2(1-u)}{um_b^2}\right]\Bigg)\Bigg]
$$
\be
-\frac{e_b\langle\bar{u}u\rangle}
{m_b^2-p^2}\exp\left(\frac{m_B^2-m_b^2}{M^2}\right)\Bigg\},
\label{Apc}
\ee
and for the parity violating amplitude
$$
A_{PV}= \frac{m_b}{f_Bm_B^2}
\Bigg \{\int_\Delta^1\frac{du}{u}
\exp\left[\frac{m_B^2}{M^2}-\frac{m_b^2-p^2(1-u)}{uM^2}\right]
\Bigg [e_u\langle\bar{u}u\rangle \varphi(u)
$$
$$
+\frac{3m_b}{4\pi^2}\left(\frac{m_b^2}{m_b^2-p^2}\right)
\Bigg\{\Bigg[e_u \left( 2u-1+\frac{p^2(1-u)}{m_b^2}
-\frac{p^2u^2}{m_b^2-p^2(1-u)}\right )
$$
$$
-e_b\left(2u-1+ \frac{p^2(1+u)}{m_b^2}-
\frac{p^2u^2}{m_b^2-p^2(1-u)}\right)\Bigg]
\frac{(m_b^2-p^2)(1-u)}{m_b^2-p^2(1-u)}
$$
\be
+e_bln\left[\frac{m_b^2-p^2(1-u)}{um_b^2}\right]
\left(2u-1+\frac{p^2}{m_b^2}\right) \Bigg\}\Bigg]
-\frac{e_b\langle\bar{u}u\rangle}
{m_b^2-p^2}\exp\left(\frac{m_B^2-m_b^2}{M^2}\right)
\Bigg \}
\label{Apv}
\ee
where $M^2$ is the Borel parameter.

In order to compactify these expressions,
the dispersion integrals for
the loop contributions  are written in terms of a new integration
variable $u=(m_b^2-p^2)/(s-p^2)$,
so that substitution of
$\Delta=(m_b^2-p^2)/(s_0-p^2)$ instead of the lower limit
in all integrals in (\ref{Apc}),(\ref{Apv})
is equivalent to subtraction of higher state contributions.
Note that the results of our calculation are, in general, the
two form factors $A_{PC,PV}(p^2)$ determining
$p^2$ dependence of the matrix element
(\ref{brhogamma}). For the problem under investigation we need only
the values of these form factors at $p^2=m_\rho^2$ or at $p^2=m^2_{K^*}$.
The sum rules are valid in a much wider region of timelike variable $p^2$ ,
parametrically up to $O(m_b^2-O$(GeV$^2$)) and practically up
to $p^2$=15 GeV$^2$ (see analysis of analogous sum rules for
heavy-to-light form factors in \cite{ali3,BBKR,BKR}).

\vspace*{2.0cm}

5. We proceed now with the numerical analysis.
The important point is that all parameters entering the l.h.s
of eqs. (\ref{Apc}), (\ref{Apv}) are known
since they also enter various other QCD sum rules; the
universality of nonperturbative inputs is
of course the main advantage
of this approach.
The value of magnetic susceptibility was determined
several times \cite{belkog},\cite{PasWil}
with essentially  the same result $\chi=-4.4 $GeV$^{-2}$ at
the normalization scale of 1 GeV. We use
$\langle\bar{u}u\rangle= - $(240 \mbox{MeV})$^3$ at this scale.
The anomalous dimension of the  $\bar{q}\sigma_{\mu\nu} q $
operator is known \cite{ShifVys} to be (-4/27)
whereas the anomalous
dimension of the quark condensate operator is (4/9) resulting in
$\chi=-3.4$ GeV$^{-2}$ and $\langle\bar{u}u\rangle= - $(260 \mbox{MeV})$^3$
at the scale $\mu_b \sim \sqrt{m_B^2-m_b^2}$ inherent to the
correlation function (\ref{corr}).
The comparatively large numerical value of the parameter $\chi$
justifies the use of leading twist approximation.
The relatively high scale $\mu_b $
is an additional argument on favor of the
asymptotic form (\ref{as}) throughout our numerical calculation.

The values of all parameters
corresponding to the $B$-channel are known from
two-point sum rules for the b-quark current:
$f_B=140~ \mbox{MeV}$, $m_b =4.7 ~\mbox{GeV}$, $s_0 = 35 ~\mbox{GeV}^2$.
(see e.g. \cite{BBKR,BKR}).
Note that the $B$-meson decay constant is taken
without $O(\alpha_s)$
corrections since they are also not included in the sum rules
(\ref{Apc}) and (\ref{Apv}) where they partially cancel
those of $f_B$. With these parameters, the amplitudes
$A_{PC}$, $A_{PV}$ are calculated at $p^2=m_\rho^2$
as a function of the Borel parameter.
The predictions of the
sum rules are very stable in a rather wide region of Borel
parameter and vary by at most 5\% with
the changes of $m_b, s_0, f_B$ within the intervals
allowed by the two-point sum rule for $f_B$.
In analogy with other sum rules we isolate a certain
interval  of Borel parameter which we determine as
$8<M^2<12 GeV^2$ . Here our
estimate of twist four operators
gives contributions below the 5\% level and higher states according to the
quark-hadron duality model contribute
less than 30\%. Our final prediction for the amplitudes defining
the matrix element (\ref{brhogamma}) for the WA decay
$B ^- \rightarrow \rho^- \gamma $ is :
\be
\label{value}
A_{PC}( B ^- \rightarrow \rho^- \gamma )
=1.1 \cdot 10^{-2}\mbox{GeV}^{-1} ,~ ~
A_{PV}(B ^- \rightarrow \rho^- \gamma)
=0.85\cdot 10^{-2} \mbox{GeV}^{-1} ~.
\label{Bminus}
\ee
with negligible variation within the interval of $M^2$ specified
above. The corresponding prediction for the amplitude
of the neutral mode $\bar{B} ^0 \rightarrow \rho^0 \gamma $
is easily
obtained by replacing $u \rightarrow d$ in the sum rules.
The prediction
in the same Borel window is :
\be
A_{PC}( \bar{B} ^0 \rightarrow \rho^0 \gamma )
=-7\cdot 10^{-3} \mbox{GeV}^{-1},~ ~
A_{PV}(\bar{B} ^0 \rightarrow \rho^0 \gamma)
=-5 \cdot 10^{-3}\mbox{GeV}^{-1}    ~.
\label{Bzero}
\ee
All these predictions have a negligible variation with the Borel
parameter within the interval specified above.
As expected, an important part of the
amplitudes (\ref{Bminus}) and (\ref{Bzero}),
$(40\div50)\%$ comes from the contribution of the photon
twist-two wave function.
The contribution from the
loop diagram with photon emission from the light quark
is about the same and amounts to $(50\div60)\%$ of the total result.
The photon emission from the heavy quark
(both perturbative and nonperturbative)
is negligibly small. The strong isospin violation (or in other words,
dependence on the flavour of the spectator quark)
which is evident from the ratios of the amplitudes of charged
(\ref{Bminus}) and neutral (\ref{Bzero}) modes, is one of the
characteristic features of the WA mechanism in contrast to
the short distance penguin amplitudes which
are independent of the flavour of the spectator quark.

Concerning the accuracy of our calculation
we first notice that omitted higher order terms are indeed very small.
The higher order quark-gluon condensate
contribution, Fig. 2c, can be safely neglected since
the quark condensate contribution itself turns out to be at the level of
1\%.
Also the estimate of
twist 4 contribution allows to neglect
the diagram  Fig. 2f containing the twist 4  three-particle
photon wave function. One may even argue that this diagram is
additionally suppressed because the gluon is emitted from
the heavy quark line
(such a suppression is present in other light-cone sum rules
for the  heavy-to light transition amplitudes,
see  e.g. \cite{BKR,BBKR} ).
The higher twist terms and
nonasymptotic corrections are the main source of uncertainty
in the light-cone sum rules involving the pion wave functions.
In the present situation, the main
uncertainty comes from the range of the parameter $\chi$
which is known with 10 \% accuracy.
Our conservative estimate of the overall accuracy
is about 15-20 \%.  Furthermore, there are  $O(\alpha_s)$
corrections to the correlation function (\ref{corr}).
Presumably, the largest part originates from
perturbative gluon exchanges in the leading diagram of Fig. 2e.
Their calculation is straightforward, but beyond the goal
of our work.

It is immediate to convert eqs. (\ref{Apc}) and (\ref{Apv})
into the  sum rules for the analogous hadronic matrix elements
$
\langle 0 \mid (\bar{d}(\bar{u})L^\mu c) \mid D^{+(0)} \rangle_\gamma$
determining the radiative decays of charmed mesons
$D \rightarrow \rho \gamma $ via WA mechanism.
One has just to replace $b$ with  $c$ , B with D  in the sum rules
(\ref{Apc}), (\ref{Apv})
and substitute corresponding parameters:
$f_D=170$ MeV, $m_c =1.3$ GeV, $s_0 = 6$ GeV$^2$,
(we take them from \cite{BBKR}) $\chi=-4.0$ GeV$^{-2}$
and $\langle\bar{u}u\rangle= - $(250 \mbox{MeV})$^3$.

In the Borel parameter range $3.0<M^2<4.5$ GeV$^2$  which satisfy all
usual requirements we obtain (with small variations)
\be
A_{PC}( D ^+ \rightarrow \rho^+ \gamma )
=-1.7 \cdot 10^{-2}\mbox{GeV}^{-1},~ ~
A_{PV}(D ^+ \rightarrow \rho^+ \gamma)
= -1.5 \cdot 10^{-2}\mbox{GeV}^{-1}~.
\label{Dminus}
\ee
\be
A_{PC}( D^0 \rightarrow \rho^0 \gamma )
= +9.6 \cdot 10^{-2}\mbox{GeV}^{-1},~ ~
A_{PV}(D ^0 \rightarrow \rho^0 \gamma)
=+5.5 \cdot 10^{-2}\mbox{GeV}^{-1} ~.
\label{Dzero}
\ee
For the correlation function (\ref{corr})
with $c$ quark currents the hierarchy of contributions
drastically changes. The magnitude of
photon emission from the $c$ quark is
almost as high as from the light quark. At the same time the
relative role of the nonperturbative emission from the light
quark (i.e. of the photon wave function contribution) compared
with the perturbative emission is larger. As a result, the
amplitudes (\ref{Dminus})  and (\ref{Dzero}) represent
interferences of various important contributions and the role of
subleading terms is in general higher, especially for $D^-$ decay
where the estimate of the size of the
twist 4 term indicates  a  20\% correction.
For this particular mode a higher accuracy is required,
including the calculation of terms omitted in our present calculation.

For completeness
we list the corresponding amplitudes for the remaining
$B$ and $D$ modes which are simply connected
to the modes considered above :
\be
A(B^{-} \rightarrow K^{*-}\gamma) \simeq
A(B ^{-} \rightarrow \rho^{-} \gamma)
\label{BKg}
\ee
and
\be
A(D^{0} \rightarrow \bar K^{*0}\gamma) \simeq
A(D ^{0} \rightarrow \rho^{0} \gamma),~
A(D ^{+} \rightarrow \rho^{+} \gamma) \simeq
A(D _{s} \rightarrow \rho^{+} \gamma) .
\label{DKg}
\ee
These relations are valid for both $PC$ and $PV$ amplitudes
in the SU(3) flavour limit, i.e. if one can neglect the variation
of form factors
$A_{PC,PC}$ from $p^2=m^2_\rho$ to $p^2=m^2_{K^*}$ or, in
the case of the second relation in (\ref{DKg}), the difference between
photon emission from $s$ and $d$ quarks. From
our sum rules we are able to safely predict that
the amplitudes at the point $p^2=m^2_{K^*}$ are $O(1\%)$ ( $O(10\%)$ )
higher for $B$ ( $D$).
The WA mechanism is  absent for the modes
$B^{0} \rightarrow \bar{K}^{*0}\gamma$,
$B_s \rightarrow K^{*0}\gamma$,$ B_s \rightarrow \phi\gamma $.
and for $D^{-} \rightarrow \bar K^{*-}\gamma) $ (with the
possible exception of
 CKM suppressed contributions).
\vspace*{2.0cm}

6. With the amplitudes determining the heavy meson
annihilation matrix element (\ref{brhogamma}),
we can finally compare the
WA effects to that of the short-distance penguin mechanism
and obtain the corrections to the branching ratios.

The branching ratio corresponding to
the decay amplitude (\ref{matrel})
with the photon transition matrix element defined in
(\ref{brhogamma}) is:
\be
BR(B^- \rightarrow \rho^- \gamma )=
\frac{G^2}{64\pi}(V_{ub}V^*_{ud}a_1)^2 f_{\rho}^2m_{\rho}^2
\left(\frac{m_B^2-m_\rho^2}{m_B}\right)^3
[A_{PC}^2+A_{PV}^2]\tau_{B^-}
\label{rate}
\ee
for $B^-\rightarrow \rho^-\gamma$ and with
obvious substitutions for the other decays.
Taking for the present analysis
$a_1=1$ , $f_\rho=216 MeV$ from the leptonic width of $\rho$-meson
which we take along with all other parameters entering
(\ref{rate}) from \cite{PD} we finally obtain for the branching ratio
corresponding to the WA mechanism :
\be
\label{br}
BR(B^- \rightarrow \rho^- \gamma )\simeq 7.0 \cdot 10^{-5}GeV^2
\left (\frac{V_{ub}}{0.0035}\right)^2
\left (\frac{a_1}{1.0} \right)^2
[A_{PC}^2+A_{PV}^2]
\ee
Here uncertainties
connected with the weak interaction vertex are emphasized and our
choice for the corresponding most uncertain parameters is shown
explicitly.
We recall once again that the estimate (\ref{rate})
depends on the validity of the factorization approximation, which
in particular, allows to neglect the photon emission from
final state quarks.

In order to compare our resluts for the WA mechanism to the
short distance (SD) penguin contribution, we cast the latter into
the form of eq. (\ref{br}).
Using the results of \cite{gre3,ali3} we obtain
$$
\mid A_{PC}(B^- \rightarrow \rho^- \gamma)^{SD} \mid=
\mid A_{PV}(B^- \rightarrow \rho^- \gamma)^{SD}\mid
$$
\be
\simeq 0.12\mbox{GeV}^{-1}\left (\frac{\mid V_{td}\mid}{0.0097}\right)
\left
(\frac{0.0035}{\mid V_{ub}\mid}\right)\left (\frac{1.0}{a_1} \right)~.
\label{SDampl}
\ee
Thus, our result in eq.(\ref{value}) amounts to about a $10 \%$
correction to the SD amplitude for
$B^- \rightarrow \rho^- \gamma$ with uncertainty at the level of
50\% from the values of the CKM matrix elements. This correction
is comparable to the present accuracy of the  hadronic
matrix elements in refs. \cite{gre3,ali3}; with more refined
techniques, however, the WA contributions should presumably not
be neglected in predicting
the branching ratio
of $B^- \rightarrow \rho ^-\gamma$ and
the CP-violating asymmetries \cite{gre3}.
In comparison, the phenomenological
estimates of the long distance penguin contributions \cite{desp3,soares}
indicate that these are quite small and probably negligible.
The correction of WA
mechanism in $\bar{B}^0 \rightarrow \rho^0 \gamma$ turns out to be
much smaller, at the level of 1\%, due to relative factor
$a_2/a_1$ and due to relative smallness of the amplitudes
(\ref{Bzero}) as compared with (\ref{Bminus}).

Unlike in radiative $B$-decays, the WA effects are crucial in the
corresponding $D$-decays. The short distance penguin contributions are
completely negligible, while the long distance terms may even lead
to observable rates  \cite{burdgo,egli,BFO}:
\be
BR(D^+ \rightarrow \rho^+ \gamma )=
\frac{G^2}{64\pi}(V_{cd}V^*_{ud}a_1)^2 f_{\rho}^2m_{\rho}^2
\left(\frac{m_D^2-m_\rho^2}{m_D}\right)^3
[A_{PC}^2+A_{PV}^2]\tau_{D^-}
\label{Drate}
\ee
Using the values in
eqs.(\ref{Dminus}) and (\ref{Dzero}), we obtain
\be
BR(D^+ \rightarrow \rho^+ \gamma )\simeq 2.7 \cdot 10^{-6}, ~~
BR(D^0 \rightarrow \rho^0 \gamma )\simeq 3.1 \cdot
10^{-6} ( a_2/0.5)^2.
\label{brd}
\ee

With the relations (\ref{DKg}) we predict the branching
ratio for $D^{0} \rightarrow \bar K^{*0}\gamma$ to be
$1.5 \cdot 10^{-4}(a_2/0.5)^2$ and for
$D_s \rightarrow \rho^+ \gamma$ to be  $2.8\cdot 10^{-5}$.

We can apply the present technology to
the decays $B\rightarrow l\nu_l\gamma$ \cite{DAT};
we simply replace the $\rho$
current with the leptonic one.
With the notation of ref. \cite{bgw}, we obtain
\be
R_B^\mu = \frac1{24\pi^2 f_B^2 m_\mu^2 m_B^4}\int^{m_B^2}_0
dp^2 (m_B^2-p^2)^3 p^2 \{ A_{PC}^2(p^2) + A_{PV}^2(p^2) \}
\ee
for the ratio between the purely leptonic decay modes with
and without photon.
A problem arises because
we now need the form factor for all values of $p^2$ and not just
at the $\rho$ mass, and the sum rule method becomes unreliable for
large $p^2$ values. We can overcome the difficulty by assuming a
pole behaviour (with $B^*$ pole for $A_{PC}$ and axial $B$ pole
for $A_{PV}$) and normalizing it with our numbers
obtained from the sum rules (\ref{Apc}) and (\ref{Apv}) at
$p^2=15 GeV^2$. The phase space suppression at large values
of the lepton pair invariant mass makes the result
insensitive to the accuracy of this extrapolation.
We then obtain $R_B^\mu \simeq 19.0$ which is in
agreement with the earlier work  \cite{bgw,aes} and underlines
the importance of the radiative leptonic decays.

Finally, we should mention that the same correlation function
(\ref{corr}) may be used to derive the sum rule for the
$B^*B\gamma$ and $D^*D\gamma$ couplings. The derivation
follows  essentially the same steps as in \cite{BBKR}
where the light-cone sum rules for the $B^*B\pi$ and $D^*D\pi$
couplings have been obtained.
The detailed analysis is beyond the scope of the present
investigation and will be presented in a separate paper
\cite{Daniel&Co}.
We just mention here that our preliminary
results on the photonic $D^*$ width
combined with its pionic widths obtained in \cite{BBKR}
agree well with experimental data.

\vspace*{2cm}

7. In this paper we have calculated a part of the long-distance
contributions to exclusive radiative rare $B$ and $D$ decays.
We have shown
that the QCD sum rules with the light-cone photon wave function
allow to calculate them rather reliably.
In the case of the
decays $B \rightarrow \rho \gamma$ the long-distance contributions
considered are small, especially in the neutral mode ,
which allows to neglect them in the determination of
$V_{td}$ through that decay. We predict branching ratios for
the corresponding $D$ decays where the long-distance weak
annihilation dominates. There are other non-perturbative effects
which can be estimated in a similar fashion; work on this is in
progress.

{\bf Acknowledgements}.

We thank D. Atwood, V. Braun, B.L. Ioffe, M. Misiak,
M. Petermann, R. R\"uckl and  A. Soni
for very useful and stimulating discussions.

\end{document}